\documentclass[conference,10pt]{IEEEtran}
\usepackage{epsfig,rotating,setspace,latexsym,amsmath,epsf,amssymb,amsfonts,bm,theorem,subfigure,epstopdf}
\usepackage{cite,authblk}
\usepackage{bbm}
\usepackage{color}
\usepackage{mathtools}
\usepackage{algorithm}
\usepackage{algpseudocode}
\usepackage{verbatim}
\usepackage{xcolor}

\algrenewcommand\algorithmicforall{\textbf{foreach}}
\algrenewcommand\algorithmicindent{.8em}

\algnewcommand\algorithmicforeach{\textbf{for each}}
\algdef{S}[FOR]{ForEach}[1]{\algorithmicforeach\ #1\ \algorithmicdo}

\IEEEoverridecommandlockouts
\allowdisplaybreaks

\begin{document}

\title{Deep Learning-Based Real-Time Quality Control of Standard Video Compression for Live Streaming }

\author{
Matin Mortaheb$^{\dag}$, Mohammad A. (Amir) Khojastepour$^{*}$, Srimat T. Chakradhar$^{*}$, Sennur Ulukus$^{\dag}$ \\
\normalsize $^{\dag}$University of Maryland, College Park, MD, $^{*}$NEC Laboratories America, Princeton, NJ \\
\normalsize \emph{mortaheb@umd.edu, amir@nec-labs.com, chak@nec-labs.com, ulukus@umd.edu}
}
\maketitle

\begin{abstract}
Ensuring high-quality video content for wireless users has become increasingly vital. Nevertheless, maintaining a consistent level of video quality faces challenges due to the fluctuating encoded bitrate, primarily caused by dynamic video content, especially in live streaming scenarios. Video compression is typically employed to eliminate unnecessary redundancies within and between video frames, thereby reducing the required bandwidth for video transmission. The encoded bitrate and the quality of the compressed video depend on encoder parameters, specifically, the quantization parameter (QP). Poor choices of encoder parameters can result in reduced bandwidth efficiency and high likelihood of non-conformance. Non-conformance refers to the violation of the peak signal-to-noise ratio (PSNR) constraint for an encoded video segment. To address these issues, a real-time deep learning-based H.264 controller is proposed. This controller dynamically estimates the optimal encoder parameters based on the content of a video chunk with minimal delay. The objective is to maintain video quality in terms of PSNR above a specified threshold while minimizing the average bitrate of the compressed video. Experimental results, conducted on both QCIF dataset and a diverse range of random videos from public datasets, validate the effectiveness of this approach. Notably, it achieves improvements of up to 2.5 times in average bandwidth usage compared to the state-of-the-art adaptive bitrate video streaming, with a negligible non-conformance probability below $10^{-2}$.
\end{abstract}

\section{Introduction}
The landscape of digital communication has been profoundly reshaped by the evolution of live streaming and video transmission. This transformation has altered how we connect, communicate, and consume content. Thanks to high-speed internet and advanced technologies, individuals and organizations can now effortlessly share real-time experiences, events, and information with a global audience through live streaming. This technology knows no geographical bounds, enabling the immediate exchange of content, whether it is live concerts, sporting events, educational seminars, or personal vlogs. The primary objective is to maintain a balance between video quality and the necessary bandwidth. In certain situations, video quality requirements vary based on the service tier, while in others, these requirements change over time due to user demand. Consequently, the pursuit of video transmission technologies that can deliver seamless live streams with minimal latency and the desired quality has been paramount.

A compelling application for such technology is evident in the transmission of closed-circuit television (CCTV) video feeds. This use case is relevant in a variety of settings, ranging from providing multiple perspectives of live events like concerts and sports matches to monitoring extensive public spaces for security reasons. Typically, each video feed is required to be transmitted at a specified video quality level, often indicated by a service tier. Different video feeds may demand varying service tiers. However, a specific video feed may transition to a different service tier on demand in a relatively stable manner, triggered by significant events or suspicious activities, for example. It is important to note that higher video quality necessitates a greater bandwidth for transmission. Therefore, the goal is to minimize the encoded bitrate for each video stream to make efficient use of the available channel bandwidth. Given that the encoded video quality (e.g., in terms of PSNR) depends on the video content such as scene, action speed, etc., adjusting the encoder parameters to meet a designated video quality threshold for each video chunk while simultaneously reducing the average video bitrate is a challenging task.

In this paper, we focus on systems that adhere to industry standards, where video compression is carried out by a video encoder (VE), e.g., an H.264 encoder \cite{wiegand2003overview}. After compression, the resulting video stream is enclosed within container frames and then sent through the physical layer. Utilizing video encoders and communication systems that conform to established standards offers various benefits, including simplified deployment, compatibility with other systems, and the utilization of specialized hardware accelerations for enhanced efficiency.

Adaptive bitrate (ABR) streaming technologies are intended to dynamically adjust the average video bitrate. Nevertheless, the quality of the compressed video is contingent on the video content itself, including factors like the visual scene's complexity, color dynamics, motion within frames, and scene variations. Even though we can influence the encoded video quality by tweaking encoder parameters, the exact relation between the output quality and these parameters is unknown. This intricacy is particularly pronounced in live video streaming, making it imperative to make real-time adjustments to video encoder parameters to optimize video quality while meeting the minimum encoding video quality.

\begin{figure}[]
 \centerline{\includegraphics[width=1\linewidth]{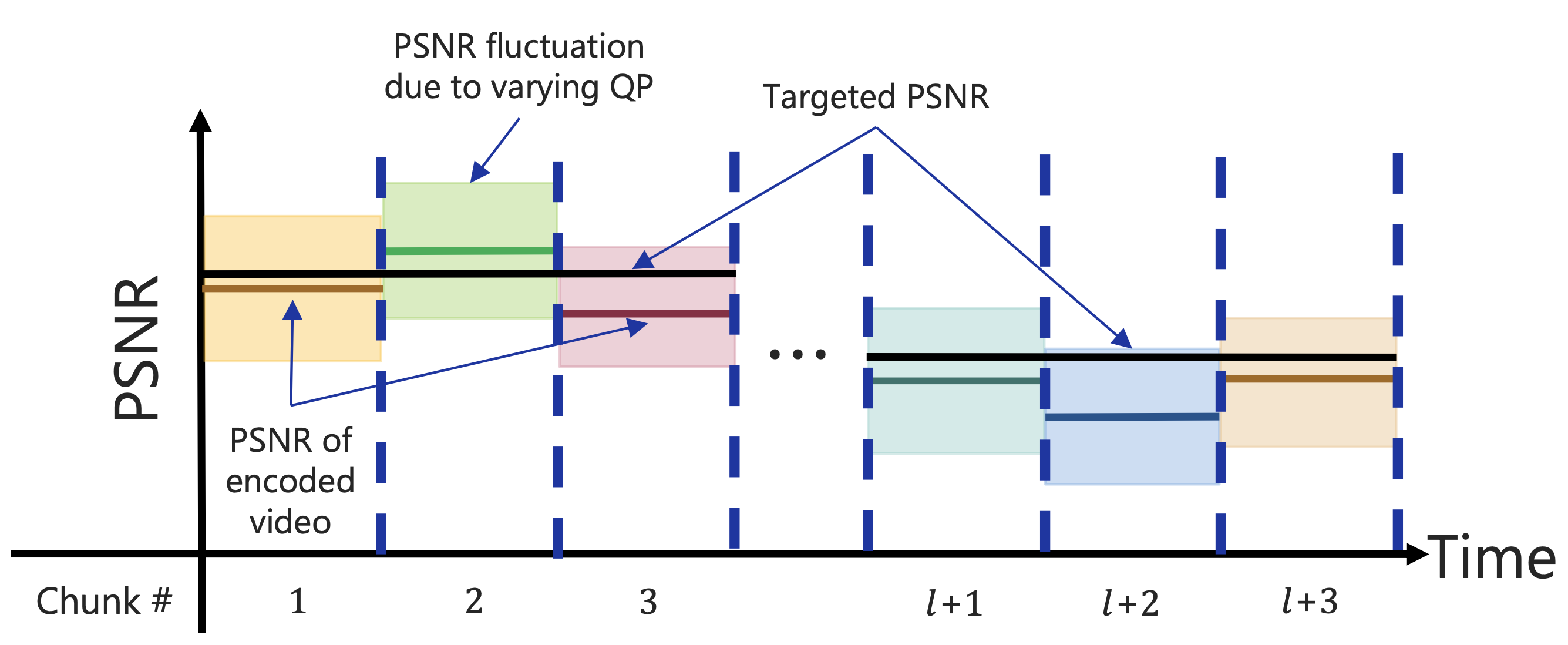}}
  \caption{Fluctuations of encoded video bitrate and available channel bitrate over chunks.}
  \label{fig:fluctuation}
\end{figure}

The visual representation in Fig. \ref{fig:fluctuation} captures the inherent fluctuations in video quality, as measured by PSNR, stemming from the dynamic and complex nature of the scene. As data is processed in chunks by the VE, the PSNR of the encoded video is assessed for each chunk individually. In Fig. \ref{fig:fluctuation}, the shaded region surrounding the solid line for each chunk showcases how the PSNR of the encoded video varies with quantization parameters $QP$ within the range of $10 < QP < 30$. It also illustrates the variation between different chunks, as well as the variability in the dynamic range itself across different chunks. This can be observed by examining the solid lines for each chunk, which represent the PSNR associated with $QP=20$ for all chunks.

To achieve a target PSNR value, indicated by black solid lines in Fig.~\ref{fig:fluctuation}, $QP$ values for each chunk must be dynamically adjusted before encoding to meet the PSNR requirement on a chunk-by-chunk basis. Fig.~\ref{fig:fluctuation} also illustrates that the target PSNR may be quasi-stationary emphasizing the fact that the user or system requirement may vary over time. 

Several well-known adaptive bitrate streaming implementations are prevalent today, such as Dynamic Adaptive Streaming over HTTP (DASH) \cite{stockhammer2011dynamic}, Apple's HTTP Live Streaming (HLS) \cite{HLS}, Microsoft's Live Smooth Streaming (Smooth) \cite{Smooth}, and Adobe's Adaptive Streaming (HDS) \cite{Adobe}. These technologies have gained significant prevalence and play crucial roles in delivering video content over the internet. The state-of-the-art in ABR transmission, represented by DASH and HLS, has unquestionably transformed internet content delivery. These systems seamlessly switch between different video quality levels, ensuring uninterrupted playback experiences. 

Despite their widespread use, both DASH and HLS protocols exhibit specific shortcomings that undermine their suitability for live streaming and real-time applications.
ABR technology in DASH and HLS relies on user-initiated and controlled selections, driven by factors like past packet receptions, device capabilities, and estimated available bitrates. This reactive approach, while effective for delivering requested video streams, falls short in actively providing the necessary quality for live streaming. Consequently, while these technologies offer valuable ABR streaming capabilities, it is crucial to consider their limitations when selecting streaming solutions for live streaming scenarios. DASH and HLS result in significant bandwidth penalties up to 250\% and an increased likelihood of non-conformance with the required PSNR constraint, affecting about 10\% to 20\% of the video chunks. These findings emphasize the need for alternative approaches.

In this paper, we present a network-aware real-time quality control (RTQC) framework. Within this framework, an RTQC controller dynamically adjusts encoding parameters in real-time to encode individual video chunks with minimal delay. This controller accurately predicts the encoded video quality, considering both the ``input video'' and the $QP$ parameter. Consequently, $QP$ is strategically selected to ensure that the encoded video quality for each chunk meets the minimum required PSNR specified by the user.

The intricate operation of the RTQC controller ensures to satisfy the required video quality for each video chunk with very high probability ($\geq 0.99$), while simultanesously minimizing the necessary bitrate to achieve such PSNR performance. The proposed technology finely customizes the video stream to accommodate user capabilities and requirements, effectively meeting the needs of all users within the network bandwidth constraints.

The rest of the paper is organized as follows. We introduce the system model in Section~\ref{sec:model}. Section~\ref{sec:RC} contains the design of RTQC scheme which is the heart of the proposed system. In Section~\ref{sec:evaluation}, we provide numerical evaluations of the proposed system model and DL-based RTQC. We conclude the paper in Section~\ref{sec:conc}.

\section{System Model}\label{sec:model}

We examine a transmitter tasked with encoding a real-time video feed, ensuring that the quality of each video segment exceeds a predefined threshold while minimizing the average bitrate of the encoded video stream. The transmitter employs a standard-compliant VE, a signal modulator and channel encoder (MCE). In this context, we adopt the H.264 codec for the VE and quantify video quality based on the PSNR of the reconstructed video segments.

Our scenario assumes a fixed number of video frames per second ($fps$) during the video capture and encoding process. Each video frame maintains a constant width ($W$) and height ($H$). With minimal delay in the VE, the video buffer preceding and following the VE exclusively contains active video frames, facilitating temporal encoding/decoding processes.

We assume that VE takes a \emph{chunk} of consecutive video frames, hence, a chunk contains one or more GOP defined as a group of pictures, i.e., video frames, over which the temporal encoding/decoding takes place. The encoded video $\Tilde{V}$ for a chunk $V$ is generated by VE with particular $H$, $W$, and $fps$. Both the PSNR $p(V, QP)$ and the bitrate of $\Tilde{V}$, i.e., the encoded bitrate $b(V, QP)$, are a function of quantization parameter $QP$ using H.264 codec. 

\begin{figure}[]
 \centerline{\includegraphics[width=1\linewidth]{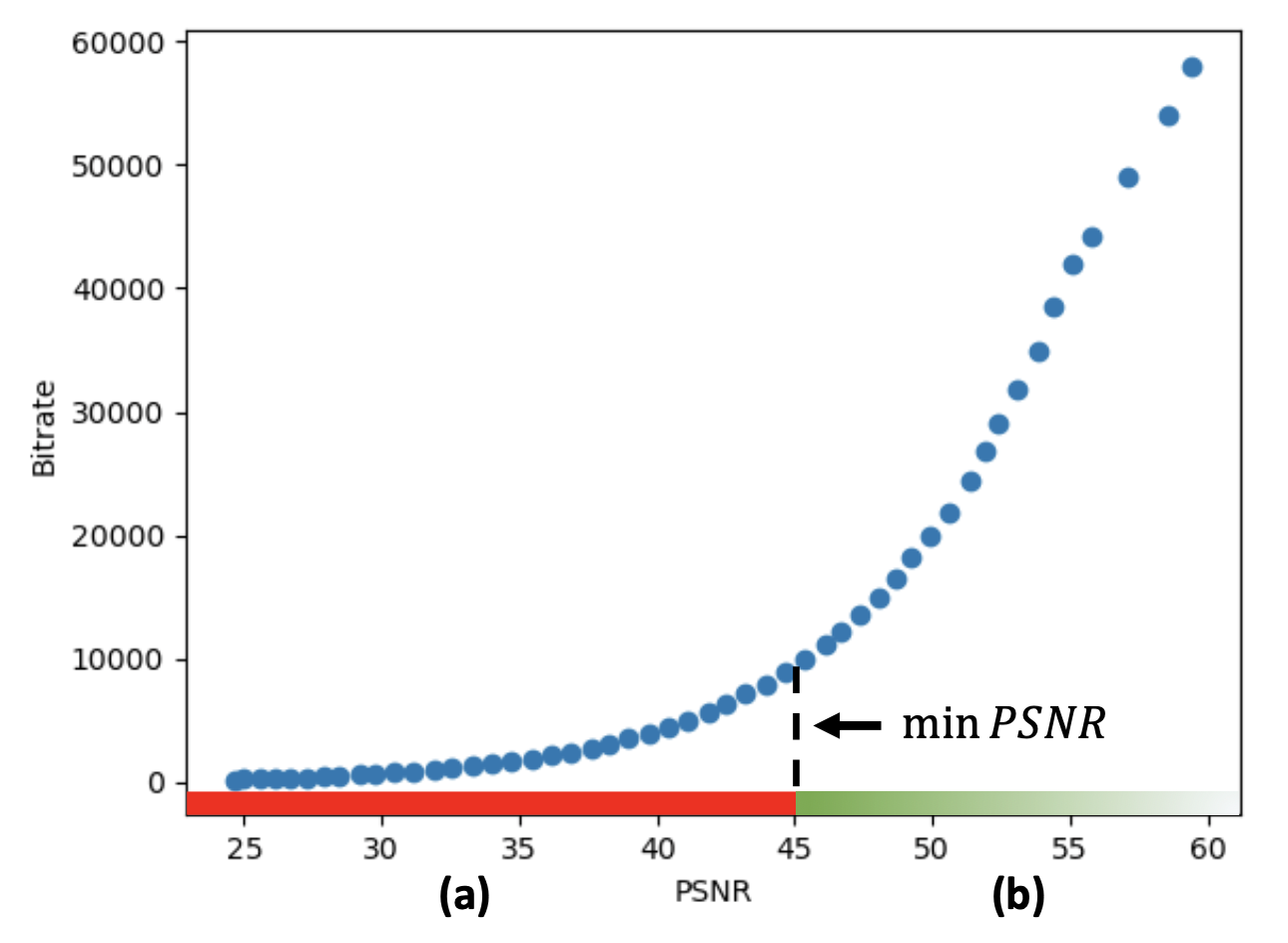}}
  \caption{Consequences of choosing wrong value for $QP$.}
  \label{fig:wrongQO_selection}
\end{figure}

The live video stream is continuously partitioned into video chunks and then each chunk is encoded using a $QP$. The encoded video is then encapsulated into \emph{container packets} which are sent through the channel. Assuming error-free reception of the container packets through the error control behavior of the physical layer, the video is then decoded. The goal is to satisfy a minimum level of the video quality of the decoded video chunks in terms of PSNR. The PSNR for each video chunk is controlled during the video compression though setting the $QP$ of a standard VE. The $QP$ which takes values from 0 to 51. The lower the $QP$ the higher the quality and essentially $QP = 0$ is a near lossless compression. 

The video quality (PSNR) and the bitrate are functions of the video content, e.g., the scene complexity and dynamics of the video frames. Hence, even though the value of $QP$ can be changed to increase or decrease the PSNR or the bitrate, the same value of $QP$ produces considerably different video quality or bitrate depending on the video content. Later in the evaluation section (Section~\ref{sec:evaluation}) we illustrate the extent of such dependency (e.g., in terms of PSNR) as a function of $QP$ for a few typical videos. The real-time nature of live streaming makes controlling the $QP$ to satisfy PSNR and minimize bitrate even more challenging. 

In live streaming, the RTQC technology should be able to identify a correct value of $QP$ that is sufficient to satisfy the constraint on video quality $p(V,QP)$ while minimizes the encoded bitrate $b(V,QP)$ in real-time. If the encoder selects a $QP$ value that is too low, the resulting encoded video quality is too high and the constraint on the encoded video quality (PSNR) can be easily satisfied. However, a low value of $QP$ leads to unnecessary high bitrate  and an inefficient utilization of the available bandwidth. Conversely, if the selected $QP$ is too high, the encoded video quality (PSNR) falls short of the target PSNR threshold (i.e., $p(V,QP) > \lambda(t)$), that results in non-conforming video chunks with respect to the system PSNR threshold $\lambda(t)$ at time $t$. 

Fig.~\ref{fig:wrongQO_selection} depicts the bitrate as a function of PSNR for a typical video. It is observed that the higher the requirement on PSNR the higher the bitrate required to satisfy the requirement. In the same figure and just above the PSNR axis, we illustrate two regions for encoded video quality with respect to a constraint on PSNR. The region (a) in the left marked with red is where we should avoid since the PSNR is not sufficient and hence the encoded video chunk is non-conforming. The region (b) in the right marked with gradient-filled green is where the PSNR is satisfied, but in order to minimize the bitrate we would like to operate as close to the PSNR threshold as possible without crossing the PSNR threshold. 

\begin{figure}[]
 \centerline{\includegraphics[width=1\linewidth]{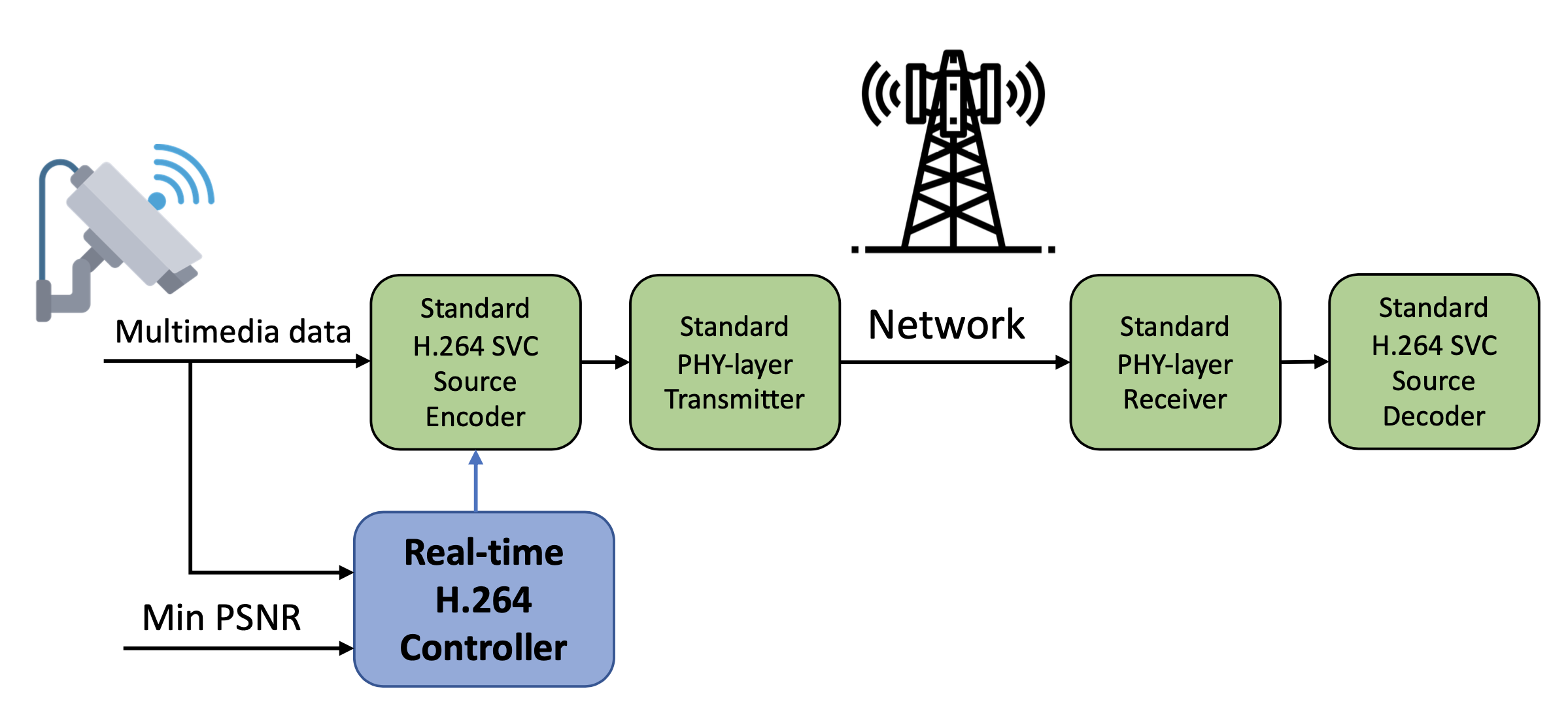}}
  \caption{System model for RTQC.}
  \label{fig:system_model}
\end{figure}

Hence, our goal is to design a deep learning-based (DL-based) network-aware RTQC which can estimate the $QP$ value in a real-time (i.e., with egligible delay) that minimizes the encoded video bitrate and satisfies a given video quality, i.e., PSNR $> \lambda(t)$. This value of $QP$ is then used by the H.264 video encoder to generate the encoded video. The complete system model can be seen in Fig.~\ref{fig:system_model}, where the proposed RTQC is highlighted with blue bounding box.

\section{Deep Learning-Based Rate Control} \label{sec:RC}

Our objective is to develop a control unit (CU) capable of accurately predicting the encoded video quality, measured in terms of PSNR. This prediction relies on both the ``input video'' and the $QP$ value. The ultimate goal is to enable the CU to intelligently select the $QP$ value, minimizing the encoded video bitrate while adhering to specified PSNR constraints. In this study, we leverage the effectiveness of a DL network. This approach has demonstrated considerable efficacy in capturing the dynamic aspects and crucial information inherent in the video, influencing the encoded PSNR of the H.264 codec across various $QP$ values.

\begin{figure}[]
 \centerline{\includegraphics[width=1\linewidth]{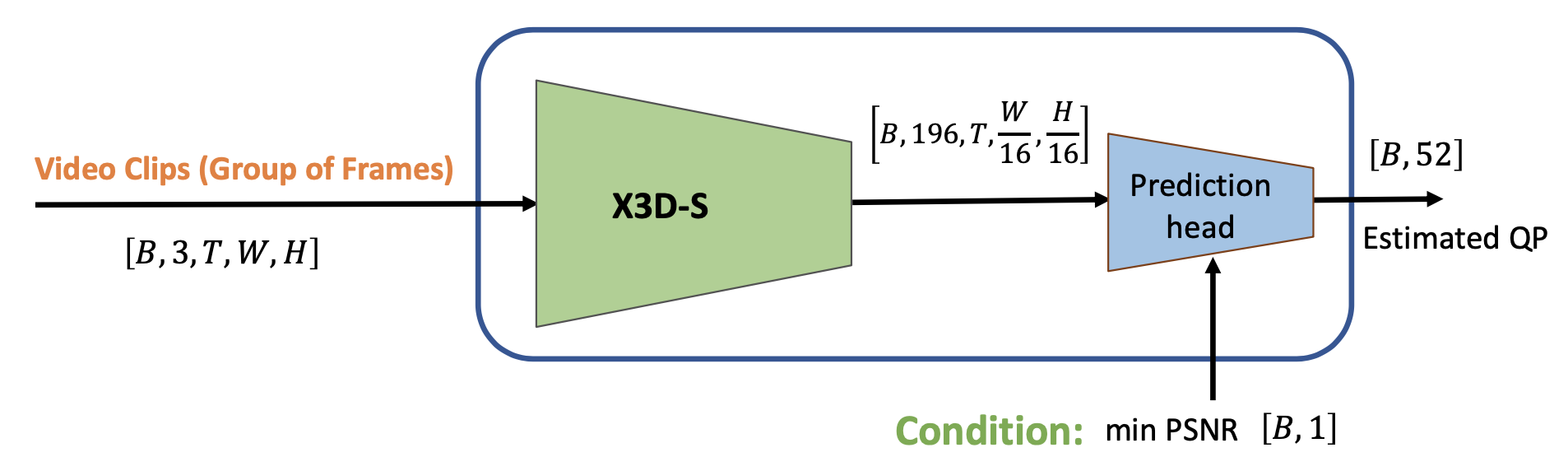}}
  \caption{Model structure for the control unit (CU).}
  \label{fig:RC_model}
\end{figure}

The architecture of the CU model is illustrated in Fig.~\ref{fig:RC_model}. In this setup, the CU is provided with an input video chunk, along with a specified minimum PSNR requirement denoted as \texttt{min PSNR}. The value of \texttt{min PSNR} is represented by $\lambda(t)$ highlighting that it may be time varying due to the change in the user demand or other system functions. The resulting output of the CU is the $QP$ value, which falls within the range of [0, 51]. This $QP$ value is subsequently utilized in the subsequent encoding stage.

The CU architecture is described as follows. We adopt X3D-S \cite{feichtenhofer2020x3d} as the foundational network for capturing video dynamics. X3D-S incorporates a diverse set of convolutional layers, significantly enhancing its ability to recognize and comprehend video content. Processing video content with dimensions $[B, 3, T, W, H]$, X3D-S produces corresponding features sized as $[B, 196, T, W/16, H/16]$, where $B$ represents the batch size, $T$ denotes the number of frames in a segment, and $W$ and $H$ signify the width and height of the video segment, respectively.

The resulting feature is further processed through a deep neural network (DNN) prediction head, which includes multiple convolutional layers, each followed by a conditional group normalization (CGN) block \cite{wu2018group}. The CGN blocks serve to normalize the output from the preceding layer. Notably, the CGN block takes a tensor of $\log_{10}(\lambda(t))$ with a size of $[B, 1]$ as a conditioning factor. Each element in this tensor represents the required minimum PSNR for each video within the batch.

The CGN processes this condition through three linear layers, each incorporating the Gelu activation function \cite{hendrycks2016gelu}, thereby transforming the tensor size to $[2B, 1]$. Subsequently, this tensor is split into two tensors of size $[B, 1]$, denoted as $\gamma$ and $\beta$. These tensors are then utilized to adjust the normalized output from the preceding layer using the formula $\gamma \times output + \beta$. This approach effectively trains the video feature to discern $QP$ values across various scenarios with varying magnitudes of $\lambda(t)$.

Our training set comprises three essential elements: a video chunk ($V$), a specified target quantization parameter ($QP_{\text{target}}$), and the corresponding minimum PSNR ($\lambda_{\text{target}}$) linked to the designated $QP_{\text{target}}$. Within our CU, we utilize $V$ as input for the X3D-S model and employ the $\lambda_{\text{target}}$ value as a conditioning factor for the CGN blocks. The outcome generated by the CU is an estimated $QP$, represented as $\widetilde{QP}$. For training, we employ the following loss function:
\begin{align}
    L = L_{CE} (\widetilde{QP},QP_{target}).
\end{align}
Here, $L_{CE}$ denotes a cross-entropy loss function. During CU training, $\lambda_{\text{target}}$ is not strictly treated as a minimum PSNR limit. Consequently, during testing, there is a possibility that the estimated $QP$ is slightly higher (usually by one or two steps) than the $QP$ value required to ensure a certain minimum PSNR. To counteract this rounding effect, we utilize slightly lower values for $QP$ than the estimated $\widetilde{QP}$ for encoding the data. This adjustment effectively prevents the occurrence of unwanted non-conforming compressed video output for the corresponding video chunk. While the resulting increase in the compressed video bitrate due to a slight decrement in $QP$ by one or two steps is tolerable, it is crucial to note that using a lower value of $QP$ than the estimated value by CU improves the packet success rate in terms of adhering to the minimum PSNR requirement. However, this approach may decrease bandwidth efficiency and increase the overall transmitted video bitrate.

\section{Performance Evaluation}\label{sec:evaluation}

In this section, we commence by assessing the variation of PSNR as a function of the $QP$ values. Our extensive studies, covering hundreds of randomly selected videos from public datasets, reveal that the coefficient of variation ($CV = \frac{\text{std}}{\text{mean}}$), defined as the ratio of the standard deviation over the mean of the PSNR, computed over 10 to 100 consecutive chunks of the same video has a mean of approximately 0.23, indicating a relatively high variability. For this study, the duration of the video chunks is set to 1 second, equivalent to 30 frames when $fps = 30$; thus, a chunk corresponds to multiple GOPs or a considerably large GOP.

\begin{figure}[]
 \centerline{\includegraphics[width=1\linewidth]{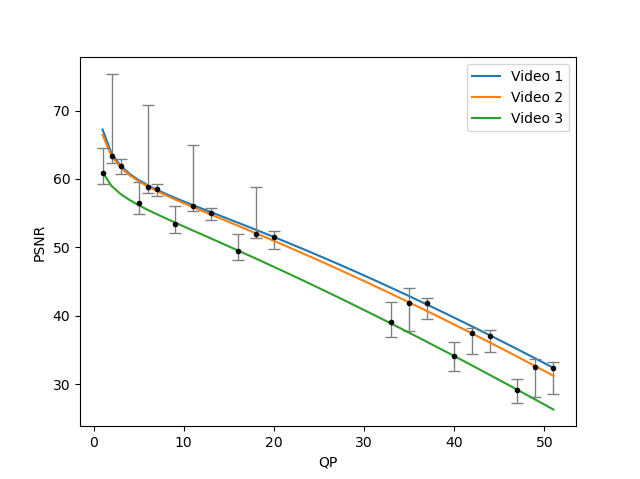}}
  \caption{Variance in PSNR of encoded video over chunks in different $QP$ values.}
  \label{fig:bitrate_variance}
\end{figure}

To visually depict the encoded PSNR fluctuations within a single video and across different videos, we select three representative videos, one from a football match, one from a live interview, and one from a concert show. Each of these three videos are partitioned into 100 one-second chunks, and each chunk is encoded using seven values of $QP$ such as $QP_1 = (1, 5, 9, 16, 33, 40, 47)$, $QP_2 = (2, 6, 11, 18, 35, 42, 49)$, and $QP_3 = (3, 7, 13, 20, 37, 44, 51)$, respectively. 

Fig.~\ref{fig:bitrate_variance} illustrates the variation in encoded video quality, measured in terms of PSNR, across each quality level. The curves denote the average PSNR for the entire video using the corresponding value of $QP$ for each of the three selected videos. The vertical bars at seven different $QP$ values in each curve mark the minimum and the maximum PSNR achieved in one of the 100 chunks for the corresponding video. We intentionally use slightly different $QP$ values for different videos so that the vertical bars do not interfere.

We observe significant PSNR variation for each $QP$ among the video chunks within each video and also across different videos. This observation underscores the importance of employing a technique, such as the DL approach presented in this paper, that can effectively understand video dynamics for each individual video chunk. This understanding enables precise adjustment of encoder parameters to meet the minimum PSNR requirement while minimizing the average encoded bitrate.

To facilitate the training of our CU, we assume that the chunk size is equivalent to one GOP with a size ($T$) of 8 frames. For videos with a frame rate of $fps = 25$, the resulting chunk size is 0.32 seconds. We make use of the QCIF dataset \cite{Dataset}, which consists of videos in an uncompressed YUV format. This dataset encompasses 25 types of videos, all in the QCIF video format, characterized by dimensions of $176$ pixels in width ($W$) and $140$ pixels in height ($H$). These videos are decomposed into uniform chunks of 8 consecutive frames, a process facilitated by the FFmpeg tool. Subsequently, we encode all the chunked videos with FFmpeg across 52 video qualities, resulting in a total of 66,768 video chunks.

For training purposes, we allocate 60,000 chunks from the complete set, reserving the remaining chunks for the testing phase. The batch size ($B$) is set to 32. In our simulation, we employ the Adam optimizer with a learning rate of $\eta = 10^{-4}$.

\begin{figure}[t]
 \centerline{\includegraphics[width=0.8\linewidth]{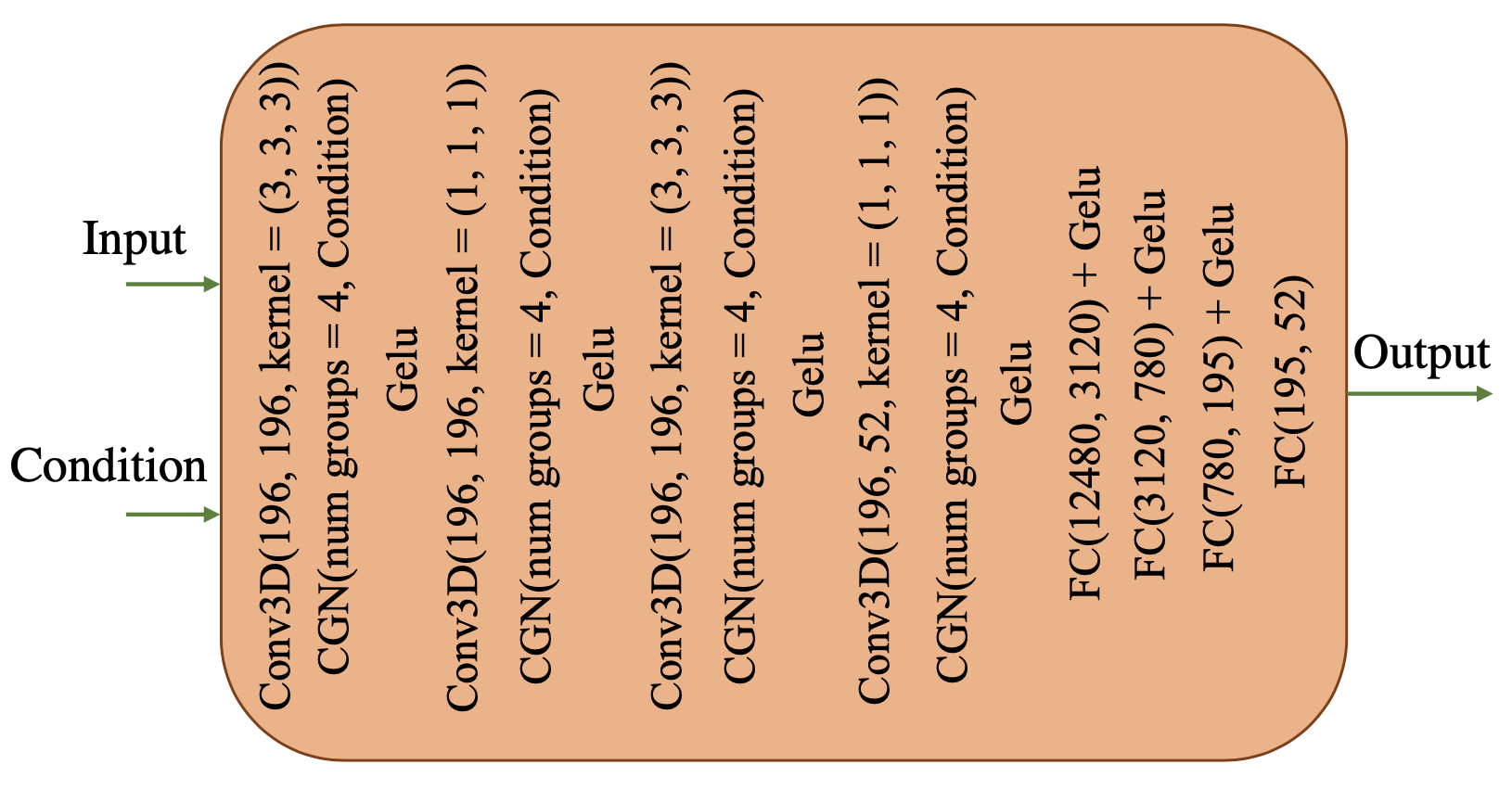}}
  \caption{Model structure for prediction head.}
  \label{fig:model_pred}
\end{figure}

\begin{figure}[t]
 \centerline{\includegraphics[width=0.6\linewidth]{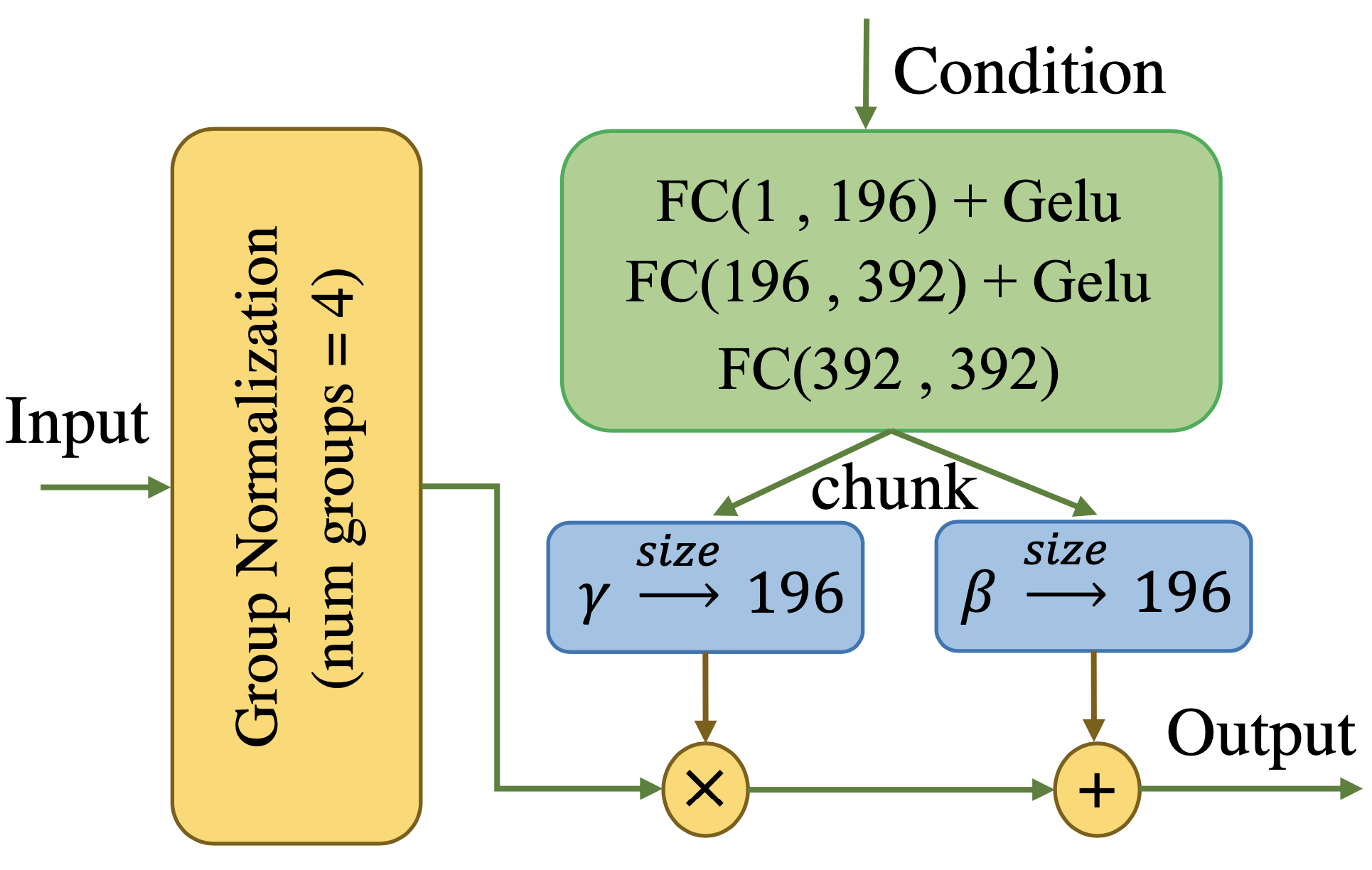}}
  \caption{Model structure for CGN.}
  \label{fig:model_CGN}
\end{figure}

We evaluate the reconstruction performance in terms of PSNR defined as the average PSNR of the video frames defined as 
\begin{align}
    \text{PSNR (dB)} = 10\log_{10}\left(\frac{\max^2}{MSE}\right).
\end{align}
The model used for CU is comprises two parts the prediction head and the CGN as shown in Fig.~\ref{fig:model_pred} and Fig.~\ref{fig:model_CGN}, respectively. 

During the training, the CU model takes an input pair comprising a video chunk (from the training-set) and its corresponding encoded PSNR and the output of the CU is compared against the input label comprising the corresponding $QP$ value for the selected training video chunk. Subsequently, we evaluate the performance of the trained CU model using the test-set. As detailed in Section~\ref{sec:RC}, we address the inherent rounding effect in our RTQC design by decrementing the output $QP$ from the DL RC model by one or two steps for the desired accuracy. While the estimated output by DL model yields accuracy of $93.8\%$, adjusting the $QP$ value one (two) step lower than the estimated output by the DL model yields an accuracy of $98.10\%$ ($99.70\%$).

\begin{figure}[t]
 \centerline{\includegraphics[width=1.\linewidth]{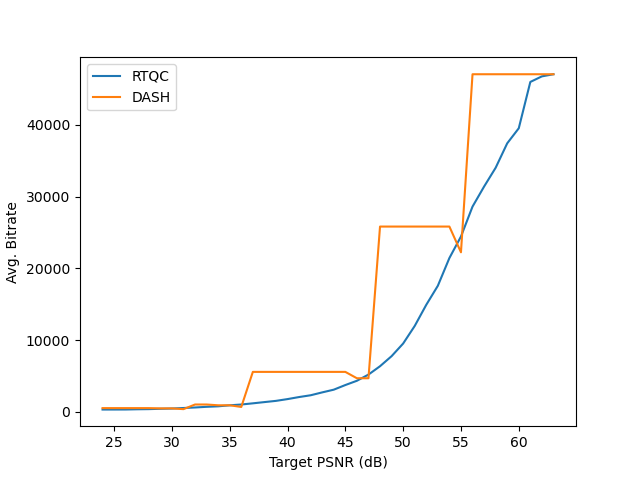}}
  \caption{Bitrate vs. target video PSNR.}
  \label{fig:PSNR_vs_ACB}
\end{figure}

Fig.~\ref{fig:PSNR_vs_ACB} provides a comparative analysis of the proposed RTQC framework and DASH algorithm in terms of average bitrate versus the reconstructed video quality (measured in trems of PSNR). In this simulation, the CU has no prior knowledge of the live video streams and for each video chunk it forecasts the proper $QP$ to satisfy the minimum PSNR and minimize the average bitrate. However, we use a very favorable situation for the DASH algorithm by assuming that DASH has prior knowledge about the average of the PSNR over the next 100 video chunk for each of its resolutions, and hence DASH picks the resolution based on this knowledge. The results demonstrate that our novel RTQC method consistently outperforms DASH by achieving considerably (up to 2.5 times) lower bitrate depending on the target PSNR value.  

The considerable improvement achieved by our novel RTQC technology lies within two key attributes. First, the ability of RTQC to dynamically adapt the encoding PSNR using 52 distinct video quality configurations in real-time, while DASH is limited to an assortment of 5 pre-established video quality levels. 
Second, the delicate configuration of $QP$ in RTQC ensures that the PSNR of the encoded frame marginally surpasses the predetermined threshold with high probability. In contrast, DASH adopts an alternative approach, striving to sustain consistent quality until the average channel bitrate surpasses the average bitrate associated with the subsequent available video quality level. To elaborate, DASH makes quality determinations grounded in average video quality observed over multiple chunks (in this context we assume a non-causal information about a future video segment of duration 10 seconds). Hence, DASH may suffer from high probability of non-conformance with the required PSNR for each chunk due to variation of the PSNR of a segment across its chunks. 

\begin{figure}[t]
 \centerline{\includegraphics[width=1\linewidth]{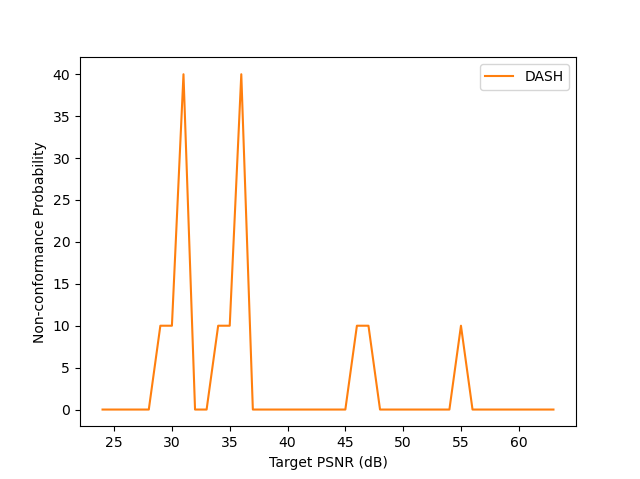}}
  \caption{Non-conformance probability vs. targeted video PSNR.}
  \label{fig:PDP_vs_ACB_fading}
\end{figure}

Fig.~\ref{fig:PDP_vs_ACB_fading} illustrates the non-conformance probability of the DASH algorithm as a function of target PSNR for DASH. It is observed that particularly at the transitions between the next video, the video PSNR is matched with the average PSNR over multiple chunks, and hence, there is considerable chance that DASH suffers from the non-conformance probability of up to 40\%. However, for the channel qualities which fall between the average bitrates of two available DASH resolutions, the non-conformance probability remains very small and close to zero. The non-conformance probability of the RTQC is omitted from Fig.~\ref{fig:PDP_vs_ACB_fading} as it is below $10^{-2}$ and does not add valuable illustration in this figure.

We define bandwidth efficiency between 0 to 1 (0 being the worst and 1 being the best) which measures how close an algorithm performs in terms of average encoded bitrate comparison to the optimal solution. Let $b_{opt}(V)$ denote the optimal bitrate for a video chunk $V$. The bandwidth efficiency is then defined as an average of the bandwidth efficiency $BW_{eff}(V)$ of the chunk $V$ defined as
\begin{align}
    BW_{eff}(V) = 1 -\frac{\max(0, b(V, \Tilde{QP}) - b_{opt}(V) )}{b(V, \Tilde{QP})} 
\end{align}
where $b(V, \Tilde{QP})$ denotes the bitrate of the algorithm for the compressed video chunk $V$ using the $\widetilde{QP}$ value for quantization parameter $QP$. The value 0 indicates the worst and 1 denotes the best bandwidth efficiency.

\begin{figure}[]
 \centerline{\includegraphics[width=1\linewidth]{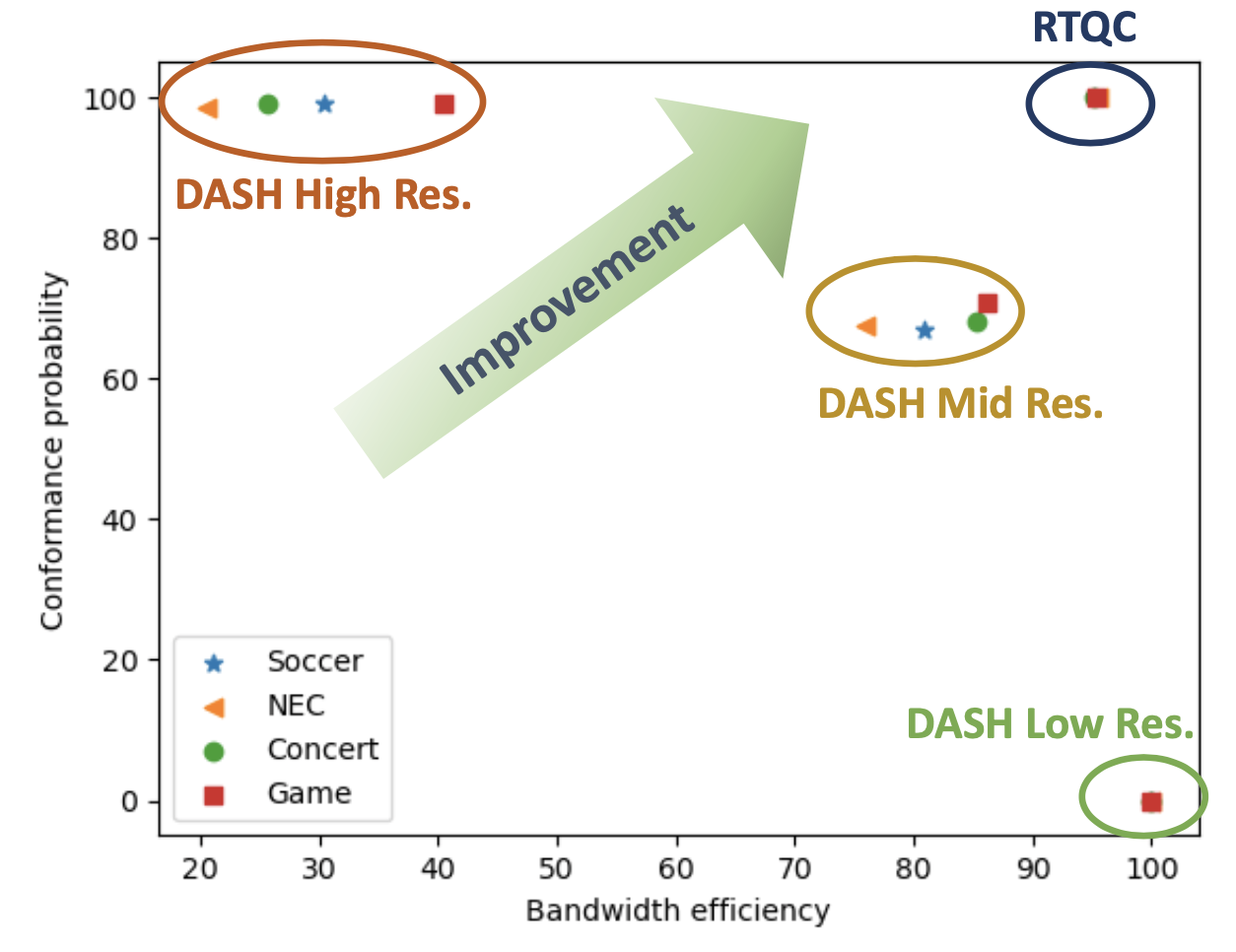}}
  \caption{Conformance probability vs. channel bandwidth efficiency.}
  \label{fig:PSR_vs_ECBE}
\end{figure}

The interplay between conformance probability and bandwidth efficiency is depicted in Fig.~\ref{fig:PSR_vs_ECBE}. In this experiment, we select four random internet videos and individually encode them at three distinct DASH resolutions: high, medium, and low. For all videos, we assume the target PSNR is equivalent to the average PSNR of the corresponding video.

Employing a high-resolution DASH configuration results in nearly 100\% conformance probability because the encoded video PSNR most likely satisfies the target PSNR. However, this approach demands high bitrate and hence it achieves only about 28\% bandwidth efficiency. On the other hand, using a lower resolution DASH encoding causes the encoded video PSNR to fall short of the target PSNR with high probability resulting in very low conformance probability but maximizing bandwidth efficiency. With mid-resolution DASH encoding, the conformance approaches 68\% due to the closely matched encoded video quality with the target PSNR. Meanwhile, this approach achieves approximately 75\% bandwidth efficiency. This phenomenon by itself reveals the fundamental trade-off between the bandwidth efficiency and packet conformance.

In contrast, the RTQC outperforms all DASH resolution scenarios in both aspects of the conformance probability and bandwidth efficiency. This superiority is due to the precise selection of $QP$ to ensure the target PSNR for the encoded video while minimizing the average bitrate, leading to nearly 98.7\% packet success rate and 99.1\% bandwidth efficiency.

\section{Conclusions}\label{sec:conc}

We proposed a DL-based real-time quality control scheme that works with existing standard video compression schemes for live-streaming of video content. The proposed DL-based real-time H.264 controller offers a promising solution by dynamically adjusting encoder parameter, namely, $QP$ values. For encoding the live stream feeds, our proposed scheme results in reconstructed video quality that conforms with target PSNR while achieving significant reduction in required bitrate in comparison to the state-of-the-art schemes such as DASH. 

\bibliographystyle{unsrt}
\bibliography{reference}
\end{document}